\documentclass[aps,pre,twocolumn,showpacs,superscriptaddress,amsmath,amssymb,longbibliography]{revtex4-2}

\usepackage[normalem]{ulem}
\usepackage{graphicx}
\usepackage{color,soul}
\usepackage{float}

\usepackage{mathtools}
\usepackage{hyperref}

\usepackage{xcolor}

\makeatletter 
\def\p@figure{\color{blue}} 
\def\p@equation{\color{blue}}
\def\p@table{\color{blue}}
\def\p@bibliography{\color{blue}} 
\makeatother

\graphicspath{ {./Figures/} }

\begin{document}

\title{Minimal model for active particles confined in a two-state micropattern}

 \author{Francisco M. R. Safara}
  \affiliation{Centro de F\'isica Te\'orica e Computacional, Faculdade de Ci\^encias, Universidade de Lisboa, 1749-016 Lisboa, Portugal }
  \author{Hygor P. M. Melo}
  \affiliation{Centro de F\'isica Te\'orica e Computacional, Faculdade de Ci\^encias, Universidade de Lisboa, 1749-016 Lisboa, Portugal }
  \author{Margarida M. Telo da Gama}
  \affiliation{Centro de F\'isica Te\'orica e Computacional, Faculdade de Ci\^encias, Universidade de Lisboa, 1749-016 Lisboa, Portugal }
  \affiliation{Departamento de F\'isica, Faculdade de Ci\^encias, Universidade de Lisboa, 1749-016 Lisboa, Portugal}
  \author{Nuno A. M. Ara\'ujo}
  \affiliation{Centro de F\'isica Te\'orica e Computacional, Faculdade de Ci\^encias, Universidade de Lisboa, 1749-016 Lisboa, Portugal }
  \affiliation{Departamento de F\'isica, Faculdade de Ci\^encias, Universidade de Lisboa, 1749-016 Lisboa, Portugal}

\begin{abstract}
We propose a minimal model, based on active Brownian particles, for the dynamics of cells confined in a two-state micropattern, composed of two rectangular boxes connected by a bridge, and investigate the transition statistics.  A transition between boxes occurs when the active particle crosses the center of the bridge, and the time between subsequent transitions is the dwell time. By assuming that the rotational diffusion time $\tau$ is a function of the position, the main features of the transition statistics observed experimentally are recovered. $\tau$ controls the transition from a ballistic regime at short time scales to a diffusive regime at long time scales, with an effective diffusion coefficient proportional to $\tau$. For small values of $\tau$, the dwell time is determined by the characteristic diffusion timescale which decays with $\tau$. For large values of $\tau$, the interaction with the walls dominates and the particle stays mostly at the corners of the boxes increasing the dwell time. We find that there is an optimal $\tau$ for which the dwell time is minimal and its value can be tuned by changing the geometry of the pattern.
\end{abstract}

\maketitle

\section*{Introduction}

Cell migration is a fundamental biophysical phenomenon with implications for morphogenesis~\cite{AMAN201020}, immune response~\cite{1} and wound healing \cite{li2013collective}. Disruption of its regulation is related to several diseases such as cancer~\cite{2}. Under biological conditions, cells move confined by other cells and by the extra cellular matrix~\cite{park2015unjamming,friedl2012new}. It is reported that cell motility correlates strongly with the surrounding conditions~\cite{paul2017cancer,pinto2020cell,dias2020modeling}. For instance, it is known that a one-dimensional confinement can trigger collective migration~\cite{pages2020cell} and that changes in the density of cells in a two-dimensional substrate can lead to an increase in their motility~\cite{melo}. Also, it was reported that metastatic cancer cells can navigate a complex porous medium selecting pore shapes to find least-resistance paths to invade~\cite{green2018pore}.

The study of individual cell migration is challenging, as it involves a hierarchical biochemical and biophysical response of the cell machinery to external stimuli~\cite{exo}. A two-state micropattern consisting of two boxes connected by a bridge has been proposed as a model system to study migration under confinement~\cite{prin.cell,fink2020area,bruckner2020disentangling}. In such a pattern, the cell performs repeated stochastic transitions between the boxes and the trajectories can be used to parameterize a stochastic equation of motion to study the time between subsequent transitions,  where the noise strength is space and velocity dependent~\cite{prin.cell}. This same pattern has also been proposed to study migration under confinement of a swarm of robots~\cite{boudet2021collections}.

The time a particle takes to escape from a box with a small opening is known as the narrow escape problem 
\cite{n1,n2}. 
For passive Brownian particles several studies have shown that the survival probability, i.e., the probability for a particle to stay in the box up to time $t$ decays exponentially with time and the mean escape time diverges as the width of the opening approaches zero~\cite{holcman2014narrow}. 
For passive Brownian particles, motion is driven by thermal fluctuations. By contrast, cells are active particles that are able to self propel~\cite{micro,cells1,cells2}. When in confined spaces, active particles tend to accumulate at the boundaries and exact calculations of the mean escape time are difficult to obtain~\cite{mori2020universal, olsen2020escape, souzy2021microbial}. So far, studies of active particles focused on one dimensional problems, and on approximation schemes~\cite{angelani2014first, malakar2018steady, weiss2002some}. These results have shown that there is an optimal strength of activity for an efficient random search of a target under confinement~\cite{olsen2020escape,rupprecht2016optimal}. 

Here, we propose a minimal model to study the stochastic transition of an active Brownian particle in a two-state micropattern. We consider a single active Brownian particle in two dimensions, in a micropattern consisting of two rectangular boxes connected by a bridge. The rotational diffusion time $\tau$ is a constant in the boxes and infinite throughout the bridge (no rotational diffusion) (see Fig.~\ref{FIG_1}). A transition between boxes occurs when the active particle crosses the center of the bridge, and the time between subsequent transitions is the dwell time. We study the dwell time $t_\mathrm{d}$, as a function of $\tau$ and the geometry of the micropattern.

We find that the average dwell time $\left \langle t_\mathrm{d} \right \rangle$ is a non-monotonic function of $\tau$ with an optimal value for which $\left \langle t_\mathrm{d} \right \rangle$ is a minimum. To understand this behavior, we decompose the stochastic trajectories in two contributions, the trajectories in the box with a random escape time $t_\mathrm{b}$, and the trajectories in the bridge with a random escape time $t_\mathrm{c}$. We find that $\left \langle t_\mathrm{b} \right \rangle$ dominates the average dwell time and it exhibits a similar non-monotonic dependence on $\tau$. When $\tau \to 0$ the motion resembles that of a passive Brownian particle, with an escape time that decays with $\tau$~\cite{benichou2011intermittent,rupprecht2016optimal,redner01}, while when $\tau \to \infty$, the motion is ballistic and the escape time is very long due to the interaction of the particle with the walls. Thus, $\left \langle t_\mathrm{d} \right \rangle$ is is expected to exhibit a minimum for intermediate values of $\tau$. We show that the position of the minimum depends on the pattern geometry, paving the way to design optimal structures for the transport of active particles.

The paper is organized as follows. We first introduce the model and describe the geometry of the pattern. In the Results section we discuss the transition statistics, where we split the contribution to the dwell time from the time in the box and the time in the bridge. Finally, we draw some conclusions in the last section.

\section*{Model}

\begin{figure}[h]
\centering
\includegraphics[width=1\linewidth]{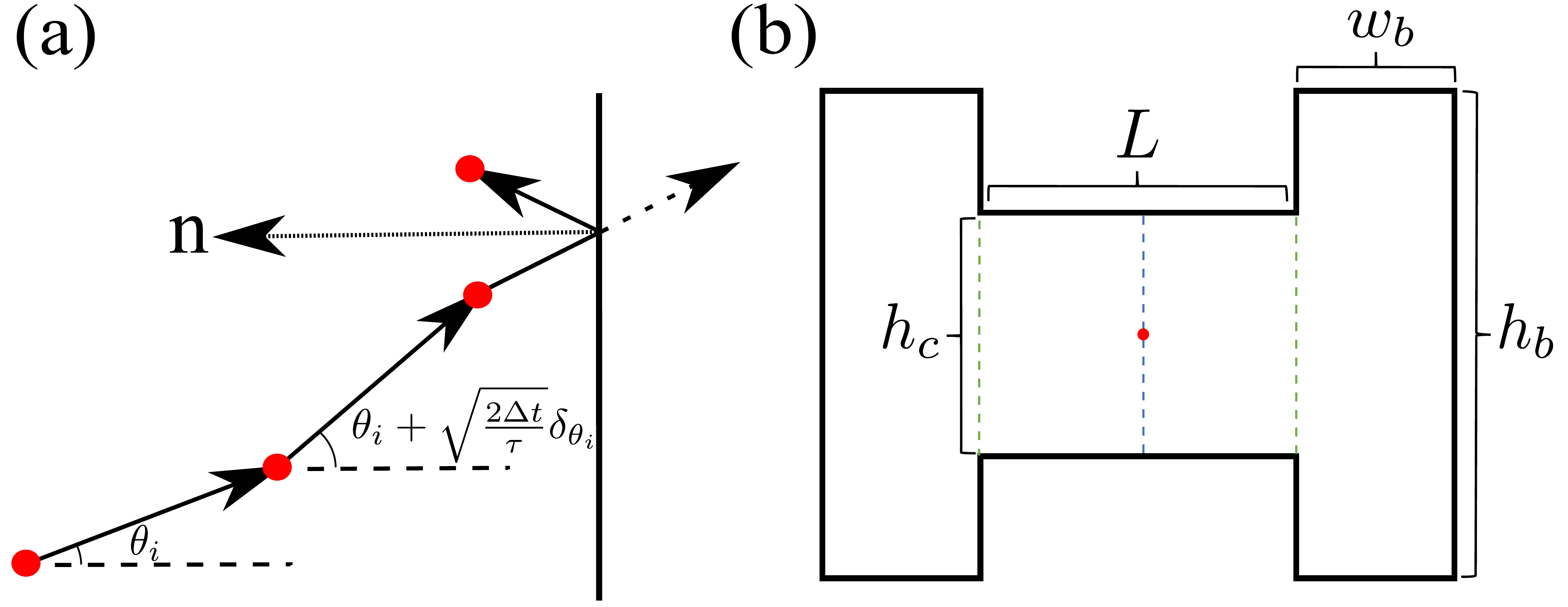} 
\caption{ \textbf{Schematic representation of the model and geometry of the pattern.} (a) Dynamics of a single  particle (red circle) described by Eqs.~(1)-(3), in one of the boxes. The random change of the direction of motion is parameterized by the rotational diffusion time $\tau=1/D_\mathrm{R}$, and the angle $\theta$ is updated as depicted in the scheme. 
In the numerical integration of Eq.~\ref{eq.1} $\delta_{\theta_i}$ is a random number taken from a Gaussian distribution with zero mean and standard deviation one.
We impose reflective boundary conditions. (b) The particle is confined to move in a pattern composed of two boxes with width $w_\mathrm{b}$ and height $h_\mathrm{b}$, connected by a bridge with width $L$ and height $h_\mathrm{c}$. The confining pattern consists of two sides, the left and the right, separated by the dashed blue line, which passes through the origin located in the geometric center of the pattern (red dot). The dashed green lines separate the boxes from the bridge.}
\label{FIG_1}
\end{figure}

On a surface, the two dimensional (2D) motion of an active Brownian particle is modeled, in the overdamped regime, by the following set of Langevin equations \cite{langevin}:

\begin{equation}
    \frac{d\theta}{dt} = \sqrt{2D_\mathrm{R}}W_{\theta},
    \label{eq.1}
\end{equation}

\begin{equation}
    \frac{dx}{dt} =v_0\cos{\theta}+\sqrt{2D_\mathrm{T}}W_{x},
    \label{eq.2}
\end{equation}

\begin{equation}
   \frac{dy}{dt} =v_0\sin{\theta}+\sqrt{2D_\mathrm{T}}W_{y},
   \label{eq.3}
\end{equation}
where $x$ and $y$ are the coordinates of the position and $\theta$ is the rotational degree of freedom. $D_\mathrm{T}$ and $D_\mathrm{R}$ are the rotational and translational diffusion coefficients, $v_0$ is the propulsion speed, and $W_{\theta}$, $W_{x}$, and $W_{y}$ represent independent white noise random variables of average zero and variance one.
For simplicity, we consider $D_\mathrm{T}=0$. Equations~(\ref{eq.1})-(\ref{eq.3}) are integrated using the Euler method \cite{prin.model}, with a time step $\Delta t$. The trajectory of the particle is resolved by recursively integrating these equations (see Fig.~\ref{FIG_1}(a)). 

Here, the motion of the particle is confined in a pattern consisting of two identical boxes connected by a bridge. 
Based on the experimental observations in Ref.~\cite{prin.cell}, we assume that cells tend to move ballistically when in the bridge, so
$\tau$ is constant in the boxes and infinite in the bridge (no rotational diffusion).

A schematic representation of the pattern is shown in Fig.~\ref{FIG_1}(b), characterized by four parameters: the height $h_\mathrm{c}$ and length $L$ of the bridge and the height $h_\mathrm{b}$ and width $w_\mathrm{b}$ of the boxes. The pattern can be divided into two sides, separated by the vertical dashed blue line.
The dashed green lines represent the separation between the boxes and the bridge. 
The collision with the walls can be simulated by considering elastic reflective boundary conditions \cite{prin.model}, as shown in Fig.~\ref{FIG_1}(a).

In what follows lengths are in units of $h_\mathrm{c}$ and time in units of $h_\mathrm{c}/v_0$. We chose $L=2$ and boxes with $h_\mathrm{b}=2$ and $w_\mathrm{b}=0.5$, all in units of $h_\mathrm{c}$, and a time step $\Delta t=0.01$ in units of $h_\mathrm{c}/v_0$.

\section*{Results}

\begin{figure}
\includegraphics[width=0.99\linewidth]{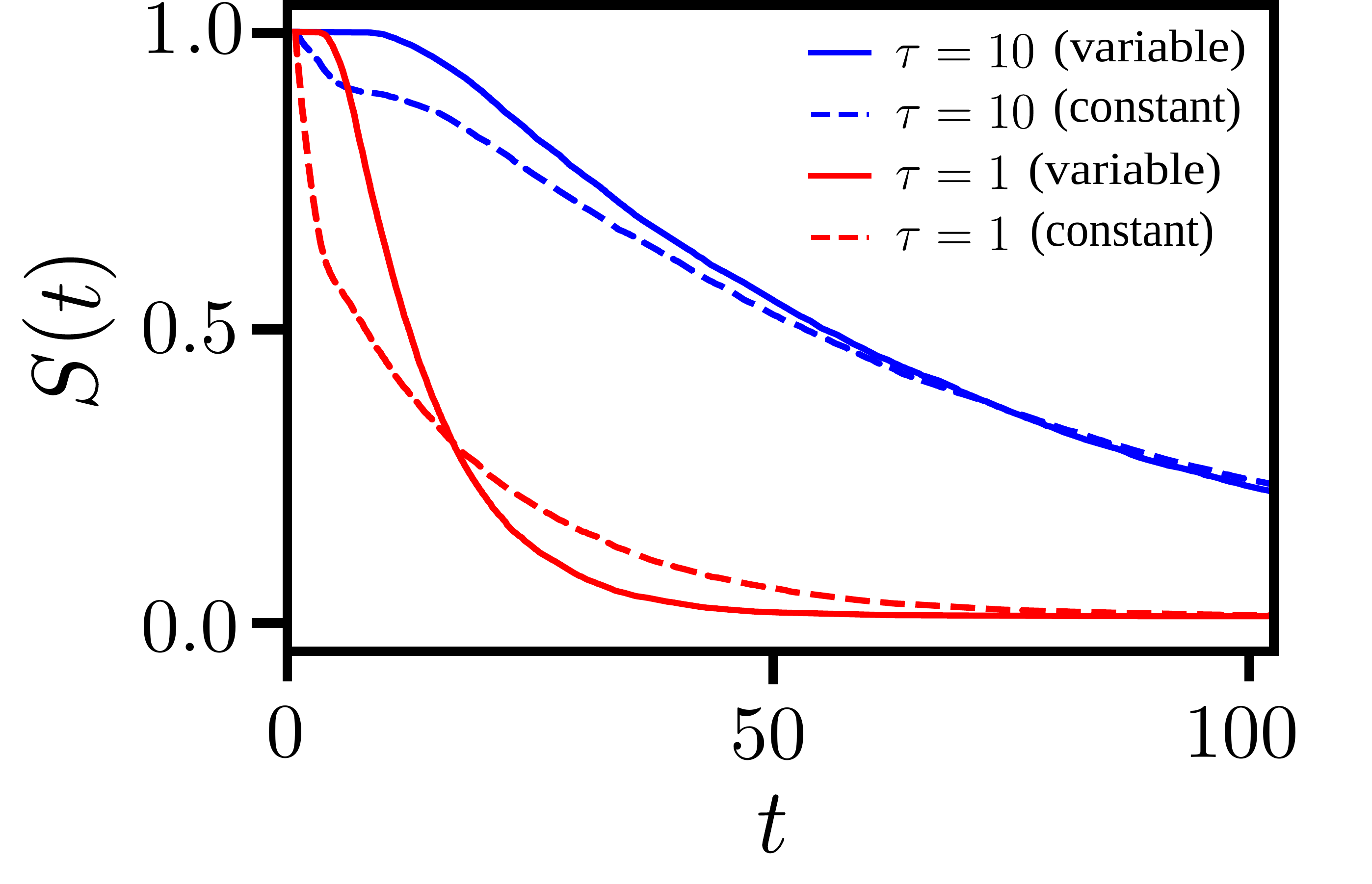}
\caption{ \textbf{Survival function.} Survival function $S(t)$ for two different values of $\tau$ in the boxes, $\tau=1$ in red and $\tau=10$ in blue, which describes the probability that the particle is in the same half of the pattern, up to time $t$. While within the dashed lines $\tau$ in the bridge is equal to that in the boxes, within the solid lines $\tau=\infty$ in the bridge. When $\tau=\infty$ there is a plateau at small values of $t$.}
\label{FIG_2}
\end{figure}

To study the dependence on the rotational diffusion time $\tau=1/D_\mathrm{R}$, we consider the survival function $S(t)$, defined as the probability that a particle remains in one side of the pattern up to time $t$. Figure~\ref{FIG_2}, shows the survival function (solid lines) for two values of $\tau$, namely, one and ten. We observe that $S(t)$ is characterized by an initial plateau and decays monotonically with time, in line with what was observed experimentally for cells in Ref.~\cite{prin.cell}. For comparison, in the same figure, we show the survival time for the same value of $\tau$ also in the bridges (dashed lines). If there is rotational diffusion in the bridge (finite $\tau$) the initial plateau disappears. Notice that, when $\tau=\infty$, once the particle is in the bridge, it will cross it, setting a minimum time to stay in one half of the pattern, which in turn leads to the plateau. By contrast, for finite $\tau$, there is a non-zero probability that, once the particle crosses the middle of the bridge it will come back due to rotational diffusion.

\begin{figure}
\includegraphics[width=0.9\linewidth]{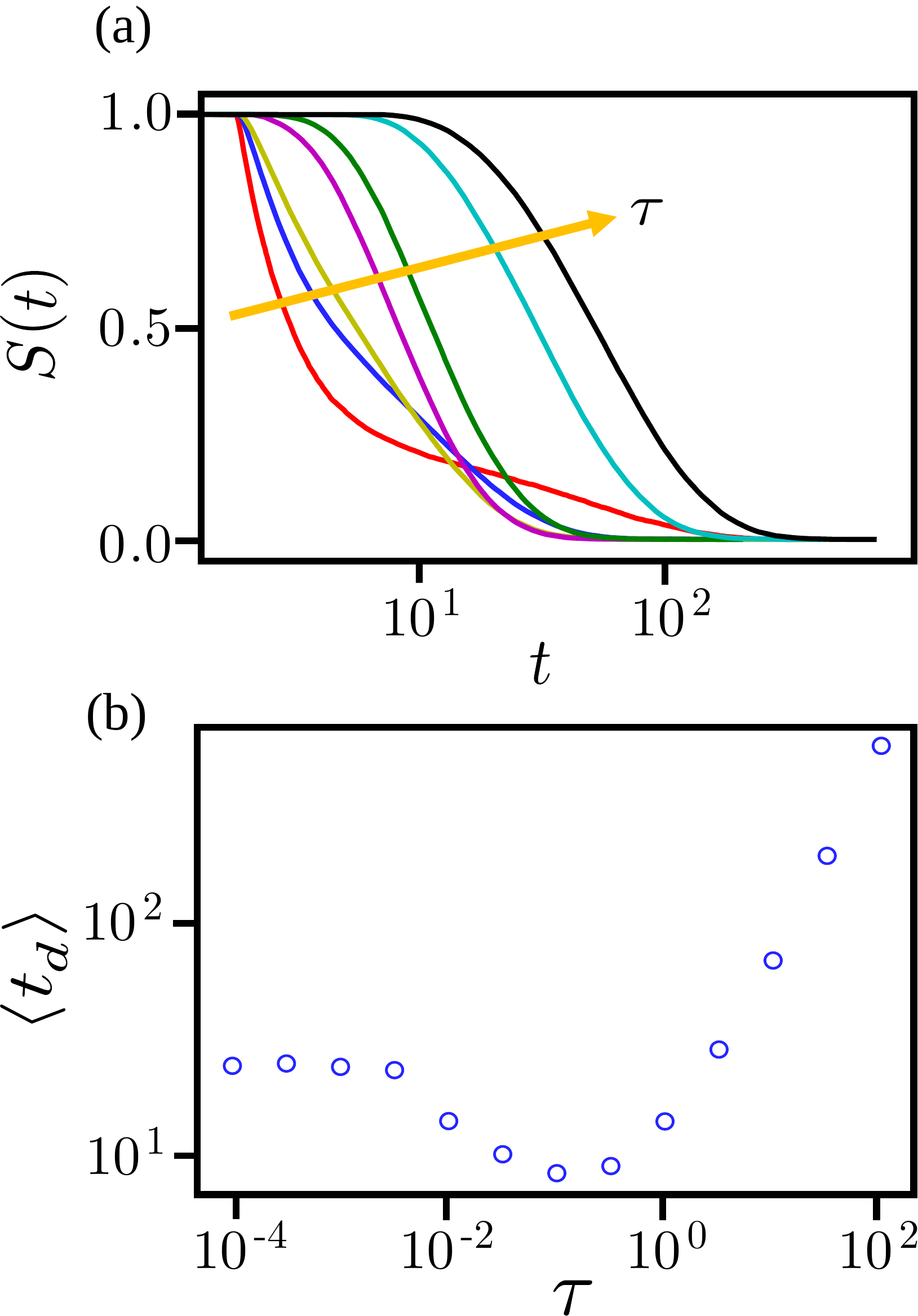}
\caption{\textbf{Statistics for different values of} $\boldsymbol{ \tau}$\textbf{.} (a) Survival function $S(t)$ for different values of $\tau$ (from left to right, following the arrow, $\tau=0.01,0.05,0.1,0.5,1,5$, and $10$). There is a change of behavior in the survival function at a value of $\tau$ of the order of $0.1$, which impacts the average dwell time $\left \langle t_\mathrm{d}  \right \rangle$. (b) $\left \langle t_\mathrm{d}  \right \rangle$ is a non-monotonic function of $\tau$, with a minimum at $\tau \approx 10^{-1}$.}
\label{FIG_3}
\end{figure}

Figure~\ref{FIG_3}(a) shows the survival function, for different values of $\tau$. All the survival functions have an initial plateau, which means that all particles take a minimum time to change sides. For the tail, we observe that there is a change in the behavior of the survival function as we increase $\tau$ and the survival function can not be rescaled by a time translation. 
Since the average dwell time $\left \langle  t_\mathrm{d} \right \rangle$ is  $\left \langle  t_\mathrm{d} \right \rangle=\int_0^\infty S(t) dt$, the change in the survival function leads to a non-monotonic $\left \langle  t_\mathrm{d} \right \rangle$ as a function of $\tau$, as shown in~Fig.~\ref{FIG_3}(b).

\begin{figure}
\includegraphics[width=1\linewidth]{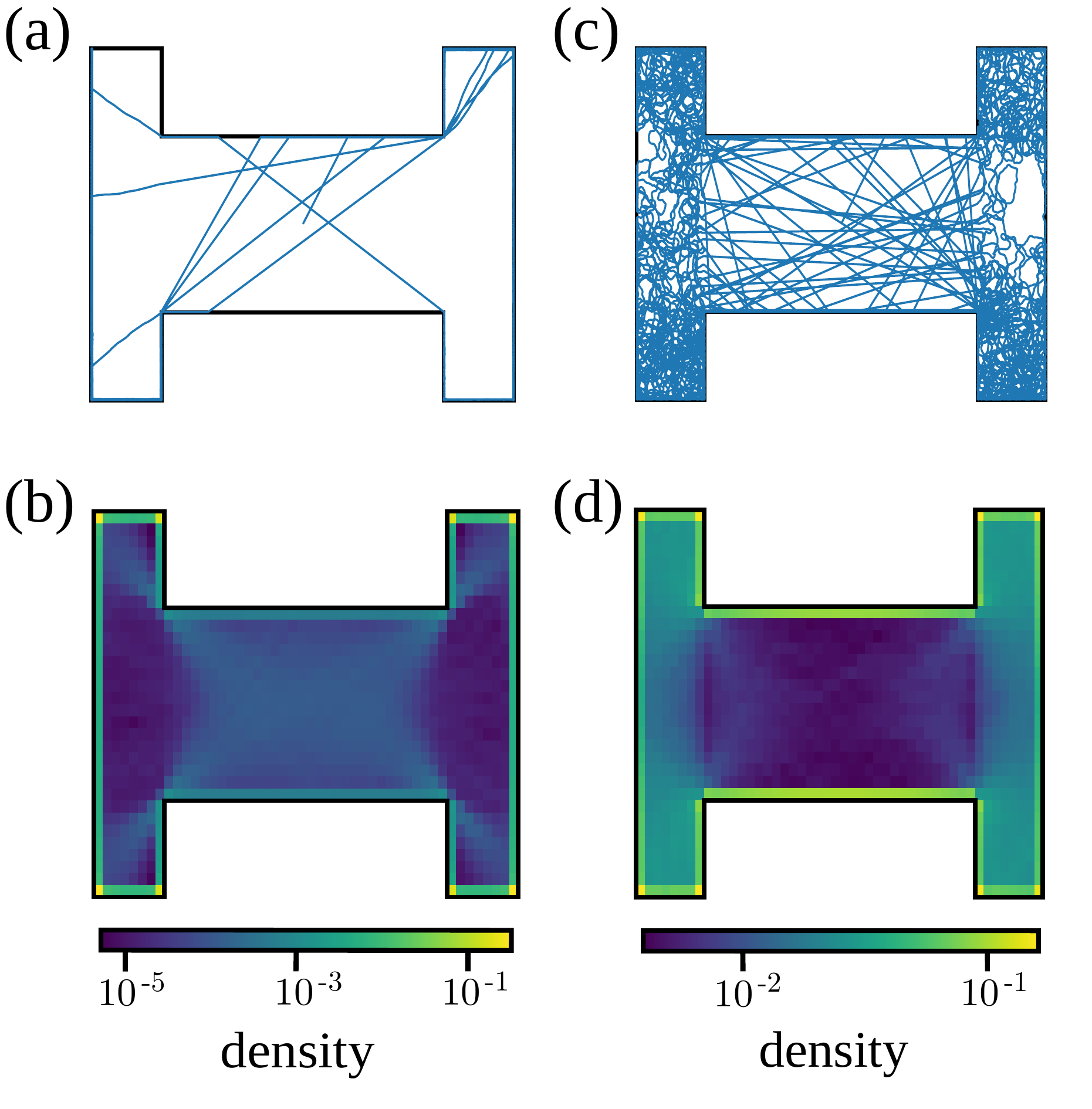}
\caption{\textbf{Particle trajectories for different values of} $\boldsymbol{ \tau}$\textbf{.} (a) Trajectory of a single active particle over a time $t=5\times10^2 $ where the time step is $\Delta t=0.01 $. The particle is confined in a pattern with $L=2$, $ w_\mathrm{b}=0.5$ and $h_\mathrm{b}=2$ and has dynamic parameters corresponding to $\tau=10$. (b) Normalized density distributions, with
$t=10^5$, and same parameters used in (a). Panels (c) and (d) show analogous results for a system with $\tau=0.1$. In the system with higher $\tau$ the particle spends more time close to the walls diffusing along them.} 
\label{FIG_4}
\end{figure}

The change in the survival function for large values of $\tau$ can be explained in the following way. When an active particle hits an obstacle such as a planar wall, the propelling force can be decomposed into two terms: one tangential and one normal to the wall. The tangential component leads to sliding along the wall while the normal component is compensated by the steric wall-particle interaction \cite{conff}. The particle will remain on the wall for a certain time, until the self-propelled direction changes. This occurs on a time scale set by $\tau$.   
In Fig.~\ref{FIG_4}(a) we plot the trajectory of a particle starting with a random angle over a time $5\times10^2$, with $\tau=10$. The particle tends to move close to the walls. This can be confirmed by analyzing Fig.~\ref{FIG_4}(b), that displays the normalized density distribution of the position of the particle. The regions where the particle spends more time are on the walls of the boxes, and particularly, an order of magnitude larger at the corners. Such an increase in time is justified by the fact that the particle gets stuck due to the large value of $\tau$. We observe that there is a depletion zone in the boxes, since the majority of events that enter the bridge tend to get stuck to its walls until reaching the boxes, exiting in an almost ballistic way.
For $\tau=0.1$ the space is explored in a more homogeneous way, as shown in Fig.~\ref{FIG_4}(c). Since this is closer to the diffusive regime, the normalized 
density distribution is homogeneous, as shown in in Fig.~\ref{FIG_4}(d).

\begin{figure*}
 \includegraphics[width=1\linewidth]{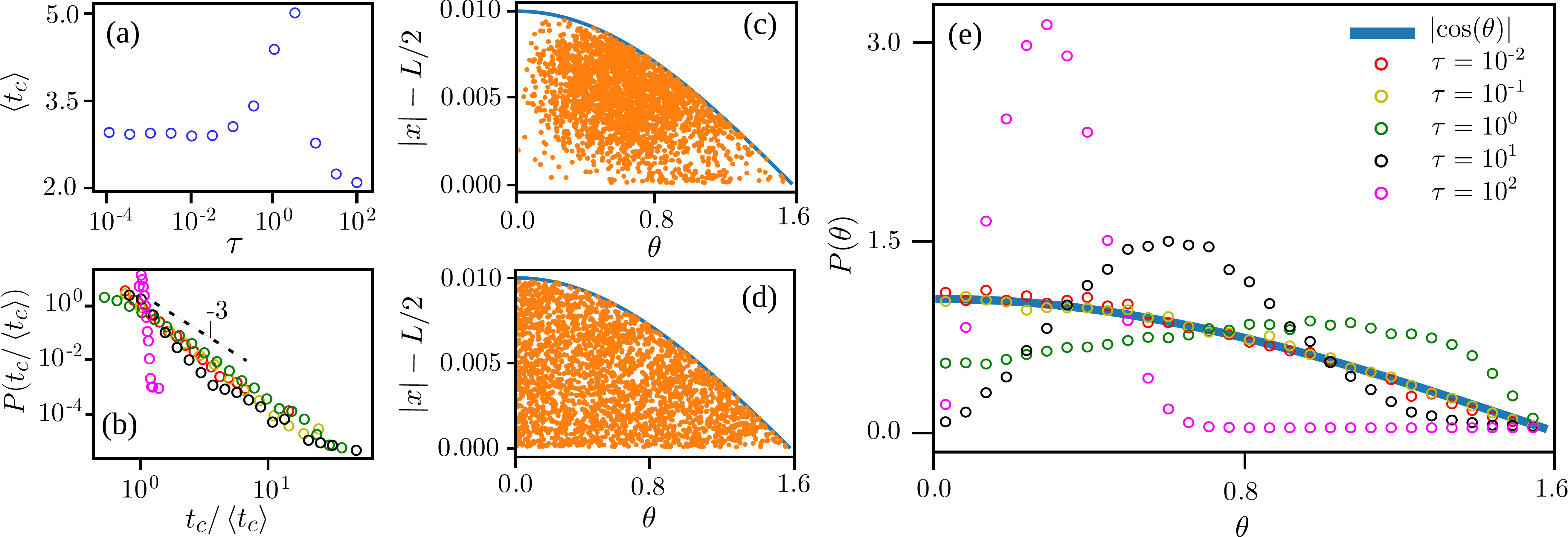} 
\caption{\textbf{Statistical properties of the time in the bridge, $\boldsymbol{t_\mathrm{c}}$, for different values of $\boldsymbol{\tau} $. }  (a) The average time $\left \langle  t_\text{c} \right \rangle$ in the bridge as a function of $\tau$, which exhibits a maximum around $\tau=3$. (b) Probability distribution of $\frac{t_\mathrm{c}}{\left \langle t_\mathrm{c} \right \rangle}$, for different values of $\tau$ labeled as in (e). For lower values of $\tau$ we find a power-law distribution with exponent close to $-3$ in line with Eq.~\ref{Eq_PowerLaw}. For large values of $\tau$ we expect deviations, as seen for $\tau=100$ where the distribution is narrow. (c) Position $\left|x\right|-L/2$ and angle $\theta$ of the particle, in the left box, before entering the bridge, for 5000 events, with $\tau=10$ (in orange). In line with Eqs.~(\ref{eq.1})-(\ref{eq.3}) we observe that this pair of values is limited by $v_0 \Delta t \cos(\theta)$ (in black). Given the symmetry of the system, it is enough to consider the angles between $0$ and $\pi/2$ and the positions of the particle in the left box. Panel (d) shows an analogous plot to (c), with $\tau=0.1$. While for small $\tau$ the distribution of ($\theta$,$\left|x\right|-L/2$) is approximately uniform in the domain of allowed values, the same does not happen for larger values of $\tau$. (e) Normalized histogram of the entrance angle $\theta$, for different values of $\tau$. The dashed line is $\left|\cos{\theta}\right|$ that corresponds to P($\theta$) for a uniform distribution P($\theta$,$\left|x\right|-L/2$), shown in (d), while for larger values of $\tau$, $P(\theta)$ exhibits a maximum that shifts to lower values of $\theta$ as $\tau$ increases.
For $\tau=100$ the distribution is peaked at one angle, which explains the narrow time distribution observed in (b).}
\label{FIG_5}
\end{figure*}

To understand the non-monotonic behavior of $\left \langle  t_\mathrm{d} \right \rangle$ as a function of $\tau$, we divide the dwell time into two contributions: the time the particle spends in the box $t_\mathrm{b}$ and the time the particle spends in the bridge $t_\mathrm{c}$. The two contributions are discussed below.

\subsection{Time in the bridge}

The total time in the bridge ($t_\mathrm{c}$) is the sum of the time the particle takes from the middle to the end of the bridge to reach the box ($t_{c_1}$) and the time it takes, after leaving the box, to reach the middle of the bridge ($t_{c_2}$). In the bridge $\tau=\infty$ ($D_\mathrm{R}=0$), the particle moves ballistically, and $t_{c_1}=\frac{L/2}{v_0 \left | \cos(\theta_1) \right | }$ and $t_{c_2}=\frac{L/2}{v_0 \left | \cos(\theta_2) \right | }$, where $\theta_1$ and $\theta_2$ are the angles with which the particle enters the bridge in each case.

The average time in the bridge, $\left \langle t_\mathrm{c} \right \rangle$, as a function of $\tau$ is plotted in Fig.~\ref{FIG_5}(a). It exhibits a maximum around $\tau=3$, a constant value $\left \langle t_\mathrm{c} \right \rangle \approx 3$ as $\tau \to 0$ and approaches $\left \langle t_\mathrm{c} \right \rangle \approx 2$ as $\tau \to \infty$. 
In Fig.~\ref{FIG_5}(b), we plot the probability distribution of $t_\mathrm{c}/\left \langle  t_\mathrm{c} \right \rangle$, for $\tau=0.01,0.1,1,10$ and $100$ (labels in Fig. ~\ref{FIG_5}(e)). We note that for the largest $\tau=100$ (pink circles) the distribution is narrow but for smaller values of $\tau$ it is a power-law distribution which we will return to it later.

In the bridge $\tau=\infty$ and, at each time step the particle moves a constant length $v_0\Delta t$. Therefore, there is a limited range of values allowed for the entrance angle $\theta$ and the distance along the x-axis from the particle to the the bridge, $ \left | x\right | - \frac{L}{2}  \leq v_0 \Delta t \left | \cos(\theta) \right | $. In Fig.~\ref{FIG_5}(c) and Fig.~\ref{FIG_5}(d) we plot the pair of values obtained by simulation, a total of 5000 events each, for two values $\tau=10$ and $\tau=0.1$, respectively. While, for the smaller $\tau$, the distribution $P(\theta,l)$, where $l=\left | x\right | - \frac{L}{2}$, is uniform, for $\tau=10$ the distribution is non-uniform. For large values of $\tau$ the particle tends to move close the walls and exits the box at the corners of the bridge (see Fig.~\ref{FIG_4}(a)), limiting the observed values of ($\theta,l$). By contrast, for smaller $\tau$ the particle enters the bridge frequently via diffusion from the interior of the boxes and, so, the walls have no major effect (see Fig.~\ref{FIG_4}(c)). 

In Fig.~\ref{FIG_5}(e), we plot the distribution of the angle $\theta$, for different values of $\tau$. If we assume a uniform distribution $P(\theta,l)$, similar to the one displayed in Fig.~\ref{FIG_5}(d), we can calculate the analytical distribution $P(\theta)$, by integrating the uniform distribution over $l$,

\begin{equation}
P(\theta) = \left | \cos(\theta) \right |.
\end{equation}
This result is in good quantitative agreement with the simulations at low values of $\tau$ (see Fig.~\ref{FIG_5}(e)). For larger values of $\tau$ the density is peaked at one angle explaining the narrow distribution of time observed for $\tau=100$ in Fig.~\ref{FIG_5}(b).

In Fig.~\ref{FIG_5}(b), we find a power-law distribution that can be rationalized as follows.  We start by assuming that the angles with which the particle enters the bridge
 are close ($\theta_1 \approx \theta_2$), and, therefore, consider a single angle $\theta$. At lower values of $\tau$ we can use Eq. (4) and calculate $P(t_\mathrm{c}/\left \langle t_\mathrm{c}  \right \rangle)$. We recall that $t_\mathrm{c}=\frac{L}{v_0 \left | \cos(\theta) \right | }$, and using the fact that $ \left | P(t_\mathrm{c}/\left \langle t_\mathrm{c}  \right \rangle) d(t_\mathrm{c}/ \left \langle t_\mathrm{c}  \right \rangle) \right |  = \left |  P(\theta) d\theta  \right |$ we obtain asymptotically

\begin{equation}
   P(t_\mathrm{c}/\left \langle t_\mathrm{c}  \right \rangle)=\frac{L^2}{t_\mathrm{c}^3 v_0^2 \sqrt{1-\frac{L^2}{t_\mathrm{c}^2 v_0^2}} } \sim t_\mathrm{c}^{-3}.
\label{Eq_PowerLaw} 
\end{equation}

Figure~\ref{FIG_5}(b), illustrates the good agreement between Eq. (5) and the simulations. The agreement is better for lower values of $\tau$ 
due to the assumption of a uniform $P(\theta,l)$. Therefore, the largest deviation from the power-law distribution is found at $\tau=100$, where the probability distribution is narrowest.

\subsection{Time in the box}

\begin{figure}
\centering
 \includegraphics[width=0.9\linewidth]{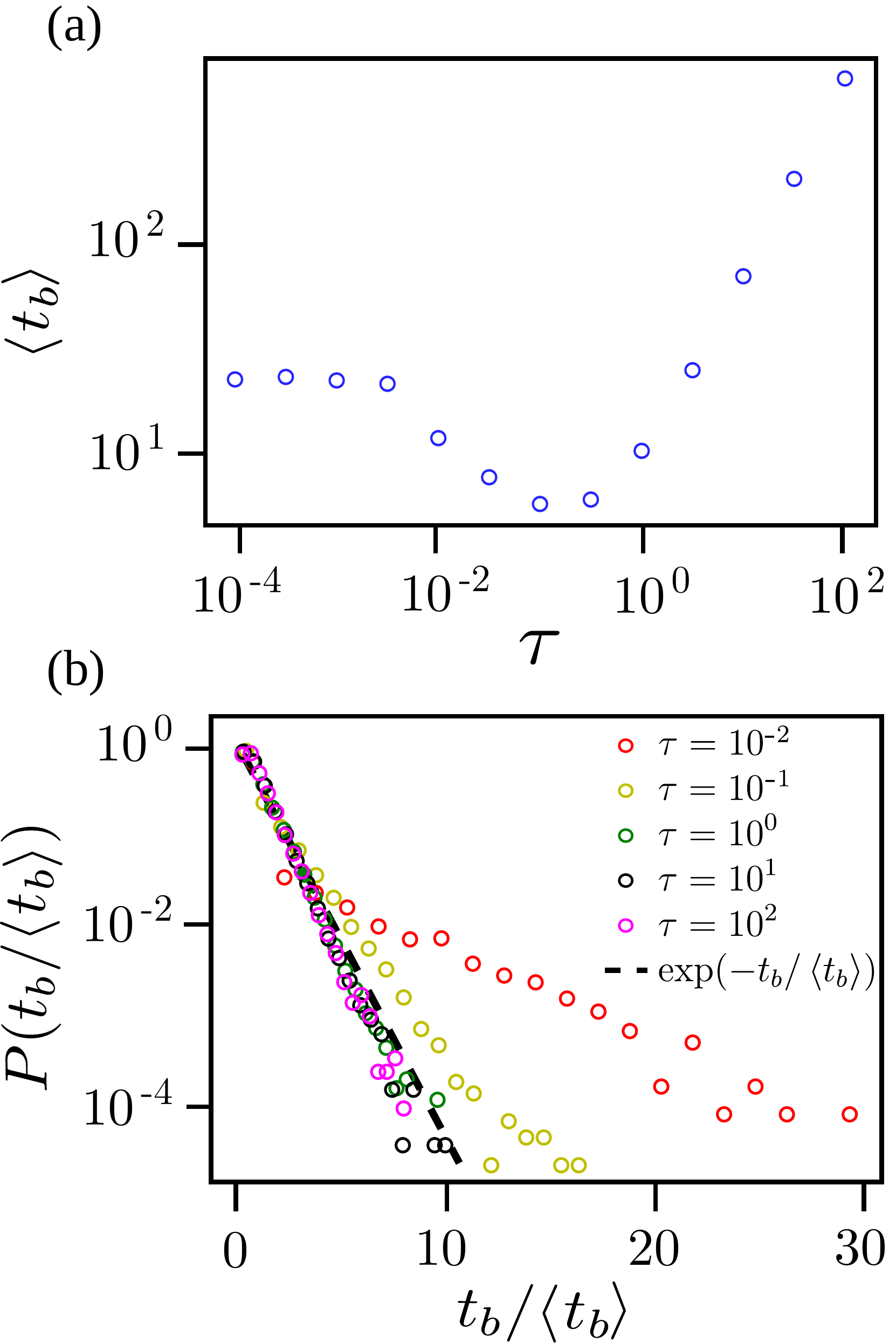} 
\caption{\textbf{Statistical properties of the time in one box, $\boldsymbol{t_\mathrm{b}}$, for different values of $\boldsymbol{\tau}$.} (a) The average time $\left \langle  t_\text{b} \right \rangle$ to escape from a box as a function of $\tau$. The behavior is very similar to the one displayed in Fig.~\ref{FIG_3}(b), that can be justified by observing that, in general, the value of $t_{d}$ is dominated by the value of $t_\mathrm{b}$. (b) Probability distribution of  $\frac{t_\mathrm{b}}{\left \langle t_\mathrm{b} \right \rangle}$, for different values of $\tau$. For large values of $\tau$ we see that the distributions collapse into a unique exponential distribution indicated by the black dashed line.}
\label{FIG_6}
\end{figure}

\begin{figure*}
\centering
\includegraphics[width=1\linewidth]{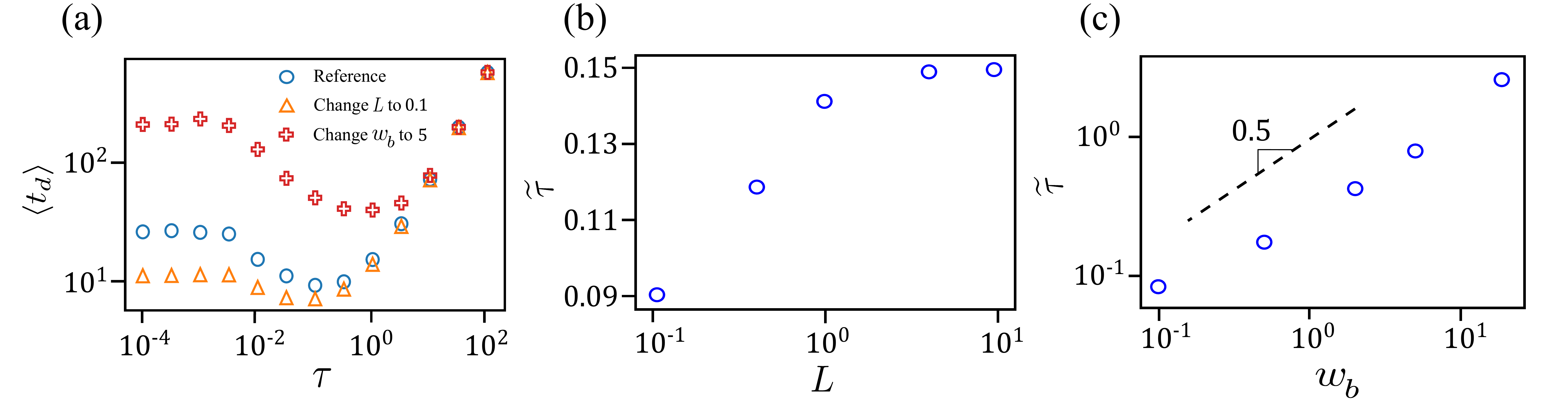} 
\caption{\textbf{Dwell time} $\boldsymbol{t_{d}}$ \textbf{as a function of} $\boldsymbol{\tau}$\textbf{, for different systems.} (a) The average dwell time $\left \langle  t_\text{d} \right \rangle$ as a function of $\tau$, for three different systems. In blue circles we display the behavior of the system with the same parameters as those used previously: $L=2$, $ w_\mathrm{b}=0.5$ and $h_\mathrm{b}=2$. In the other curves one of the parameters was changed: $L$ or $w_\mathrm{b}$. In yellow triangles $L$ is $0.1$ and in red crosses $w_\mathrm{b}$ is $5$. We observe that the value of $\tilde{\tau}$, where $\left \langle  t_\mathrm{d} \right \rangle$ is minimal, is a function of the geometry of the system. (b) Value of $\tilde{\tau}$ as a function of $L$ with  $ w_\mathrm{b}=0.5$ and $h_\mathrm{b}=2$. We observe that $\widetilde{\tau}$ increases with $L$ but saturates for $L>1$. (c) Value of $\widetilde{\tau}$ as a function of $w_\mathrm{b}$ with $L=2$, and $h_\mathrm{b}=2$. We observe that $\widetilde{\tau}$ grows as a power law of $w_\mathrm{b}$. The black dashed line represents the slope of the analytical exponent $0.5$.}
\label{FIG_7}
\end{figure*}

\begin{figure}[h]
\centering
\includegraphics[width=0.95\linewidth]{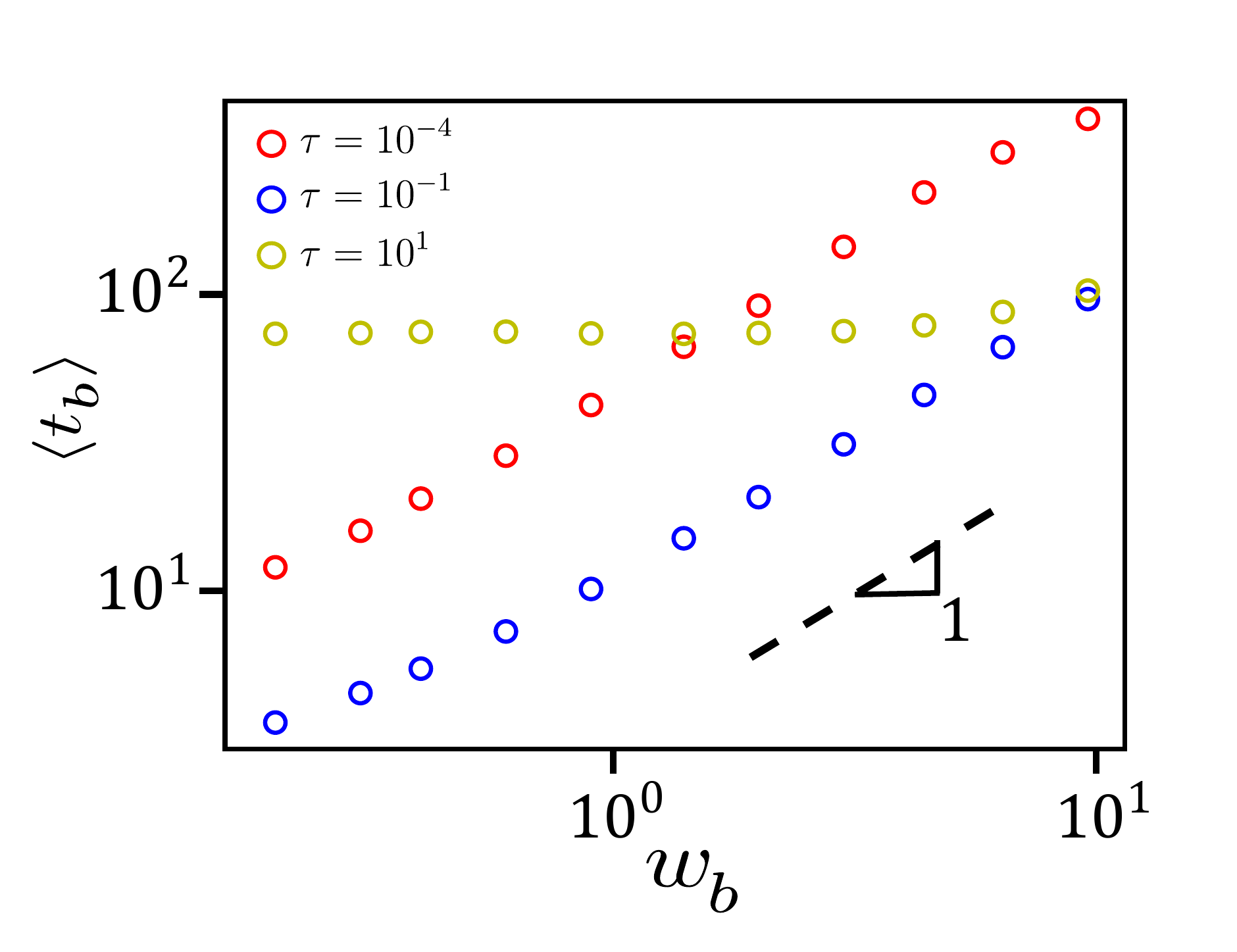} 
\caption{\textbf{Average time in the box $\boldsymbol {\left \langle t_b \right \rangle}$ as a function of the box width $\boldsymbol {w_b}$}. Average time in the box $\left \langle t_b \right \rangle$, as a function of the width of the box $w_b$ for three different values of $\tau$, with $L=2$ and $h_b=2$. For small values of $\tau$, $\left \langle t_b \right \rangle \sim w_b$, as indicated by the dashed line with slope $1$. For large values of $\tau$, $\left \langle t_b \right \rangle$ is constant.}
\label{FIG_8}
\end{figure}

In Fig.~\ref{FIG_6}(a), we plot $\left \langle t_\mathrm{b} \right \rangle$ as a function of $\tau$. In line with the dwell time (see Fig.~\ref{FIG_3}(b)),  $\left \langle t_\mathrm{b} \right \rangle$ is a non-monotonic function of $\tau$. This results from the deterministic dynamics in the bridge and the fact that for the simulated sizes of the bridge and boxes, $t_\mathrm{b}$ dominates the dwell time.

In Fig.~\ref{FIG_6}(b), we illustrate the probability distribution of $\frac{t_\mathrm{b}}{\left \langle t_\mathrm{b} \right \rangle}$, for  different values of $\tau$. For values of $\tau \geqslant 0.1$ the numerical results are approximated well by an exponential distribution. For lower values of $\tau$ the particle needs a long time to reach and explore the walls of the box and deviations from the exponential distribution are observed. This resembles what is observed for passive Brownian particles in closed domains, where the tail of the escape time follows an exponential
distribution, while at short times it is a power law, with the crossover set by the time taken by the particle to explore the size of the domain.~\cite{redner01}.

\subsection{Effect of the geometry}

The optimal value of $\tau$ depends on the geometry of the confining pattern. To explore this dependence, in Fig.~\ref{FIG_7}(a) we show the relation between $ \left \langle  t_\text{d} \right \rangle$ and $\tau$, for various geometrical parameters. We find that the position of the minimum changes while for larger values of $\tau$ the average dwell time $\left \langle t_\mathrm{d} \right \rangle$ is independent of the geometrical parameters. This is due to the long times spent by the particles at the corners of the boxes.

For each $\left \langle  t_\text{d} \right \rangle$ as a function of $\tau$ there is an optimal value $\widetilde{\tau}$ for which $ \left \langle  t_\text{d} \right \rangle$ is a minimum. Figure~\ref{FIG_7}(b) displays the relation between $\widetilde{\tau}$ and  $L$, at constant $w_\mathrm{b}$ and $h_\mathrm{b}$. Initially, $\widetilde{\tau}$ increases with $L$, but for large values of $L$ all the particles exit the bridge at the edges, and $\widetilde{\tau}$ saturates for $L>1.0$.
Figure~\ref{FIG_7}(c) illustrates similar results as a function of $w_\mathrm{b}$, other parameters fixed. In this case, 
$\widetilde{\tau}$ increases as a power of $w_\mathrm{b}$.

To calculate the exponent of this power-law we use an approximation following Ref.~\cite{rupprecht2016optimal}. Since we vary $w_\mathrm{b}$ only, the optimal value $\widetilde{\tau}$ may be obtained by minimizing $\left \langle  t_\text{b} \right \rangle$. In the diffusive regime, the main contribution for the average time in the box is $\left \langle  t_\text{b} \right \rangle \sim 1/\tau$, while for the ballistic regime $\left \langle  t_\text{b} \right \rangle \sim \tau$. Therefore, close to the minimum we use the {\it Ansatz} 
\begin{equation}
	\left \langle  t_\text{b} \right \rangle = f(w_\mathrm{b}) \tau + \frac{g(w_\mathrm{b})}{\tau}.
\label{Eq_Ansatz}
\end{equation}
Where $f(w_\mathrm{b})$ and $g(w_\mathrm{b})$ are unknown functions of $w_\mathrm{b}$ all the other parameters being held constant. The minimization condition leads to:
\begin{equation}
	\widetilde{\tau} = \sqrt{\frac{g(w_\mathrm{b})}{f(w_\mathrm{b})}}.
\label{Eq_Ansatz2}
\end{equation}
We calculated these functions numerically. In Fig.~\ref{FIG_8}, we show that for $\tau<\widetilde{\tau}$, the numerator $g(w_\mathrm{b})\sim w_\mathrm{b}$, and for $\tau>\widetilde{\tau}$, the denominator $f(w_\mathrm{b})$ is constant. Thus, $\widetilde{\tau} \sim w_\mathrm{b}^{0.5}$, in good agreement with the results of the simulations shown in Fig.~\ref{FIG_7}(c).

\section*{Conclusion}

We studied the stochastic transitions of an active particle in a confined two-dimensional pattern, composed by two boxes connected by a bridge. The particle moves with a rotational diffusion time in the boxes $\tau=1/D_\mathrm{R}$,  where $D_\mathrm{R}$ is the rotational diffusion coefficient, while in the bridge the particle moves ballistically ($\tau=\infty$). Every time the particle crosses the center of the bridge a transition occurs with a dwell time $t_\mathrm{d}$. The model is parameterized by the length scale $h_\mathrm{c}$ and time scale $h_\mathrm{c}/v_0$, where $h_\mathrm{c}$ is the bridge height and $v_0$ is the propulsion speed. Using the values reported in Ref.~\cite{prin.cell} we obtain a dwell time of the order of hours for $\tau<1.0$, in line with the values obtained experimentally.

We found that there is an optimal value of $\tau$ when the average dwell time is minimized. We split the contributions to the dwell time in the time the particle takes to escape from the box, $t_\mathrm{b}$, and the time taken to cross the bridge, $t_\mathrm{c}$. We derived an analytical expression to estimate the probability distribution of $t_\mathrm{c}$, for small values of $\tau$, and showed that, asymptotically, $P(t_\mathrm{c})\sim t_\mathrm{c}^{-3}$. For the trajectories in the box, we found that the probability distribution $P(t_\mathrm{b})$ is exponential, with a characteristic time that, like $ \left \langle  t_\mathrm{d} \right \rangle$, exhibits a minimum as a function of $\tau$. The presence of this minimum results from the balance between the diffusive and ballistic motions. Diffusion is known to be inefficient to find small targets in 2D \cite{redner01,n1,n2}. However, for active particles, the persistent motion at short times increases the time the particles spend near the walls, and as a result there is an optimal value of $\tau$ that minimizes the average dwell time.

We found that this optimal time can be tuned by changing the geometry of the pattern, such as the length of the bridge and/or the width of the boxes, which can be used to design experimental micropatterns to control the flow of cells, by increasing or decreasing the probability of the transition.

In the context of ecology, it is observed experimentally that connecting two habitats by a bridge can enhance the survival time of species moving randomly in space~\cite{pimm2019connecting, beier1998habitat, gilbert2010meta, ibagon2021contact}. We have shown that the introduction of activity may change the transition dynamics with possible implications for these ecological systems.

\begin{acknowledgments}
The authors acknowledge financial support from the Portuguese Foundation for Science and Technology (FCT) under Contracts no. PTDC/FIS-MAC/28146/2017 (LISBOA-01-0145-FEDER-028146), UIDB/00618/2020, and UIDP/00618/2020.
\end{acknowledgments}

\bibliography{bibliografia.bib}

\end{document}